\documentstyle[hx37_latex]{article}

\def\be{\begin{equation}}
\def\ee{\end{equation}}
\def\bea{\begin{eqnarray}}
\def\eea{\end{eqnarray}}
\begin{document}

\title{SURFACE-BRIGHTNESS EVOLUTION OF CLUSTER GALAXIES}

\author{David Schade }

\address{Dept. of Astronomy, University of Toronto,\\
60 St. George St., Toronto, M5S 3H8}


\maketitle\abstracts{Surface brightness evolution has been detected
in elliptical galaxies (consistent with passive evolution models
of old stellar populations) and in disk galaxies (presumably due
to enhanced star-formation rates). 
The rates of evolution in clusters and 
the field are not measurably different. In addition to this
similarity, the high-redshift populations in both environments
exhibit a ``blue-excess'' population, increased rates of star
formation, and high frequency of peculiar structure. Thus, there
are several parallels between evolving cluster and field galaxies and
the high-redshift cluster environment will be understood only
by comparison with the field population {\em at the same epoch}.}

\section{Introduction}

 In the course of programs to understand the evolution
of galaxies by focusing on morphological properties,
 two-dimensional surface photometry has been done for
samples of field galaxies\cite{s95}$^,$\cite{s1} 
and cluster galaxies\cite{s1}$^,$\cite{s2} using
$HST$, ground-based imaging, and $HST$ archival data. 
 The cluster and field galaxy populations were much more similar
at $z\sim 0.5$ than they are at the present time.

\section{Recent results on high-redshift clusters}

\subsection{Elliptical galaxies}

 Ellipticals are present in clusters at high redshift\cite{di}
and their colours are consistent with passive evolution 
models.\cite{dr2}$^,$\cite{ar}$^,$\cite{ra}$^,$\cite{ok}
The small dispersion in the colour-luminosity relation
at $z\sim 0.5$\cite{el} 
suggests they formed at substantially earlier epochs. 
 Bender, Ziegler \& Bruzual\cite{be} find evidence from the Mgb-$\sigma$
and Faber-Jackson relations for passive evolution in the $B$-band of 
$0.5\pm 0.1$ in a cluster at $z=0.37$, a result consistent with
fundamental plane work at $z=0.39$\cite{va}.

 Imaging has been used to search for evolution
in the size-luminosity relation---one projection of the
fundamental plane---of elliptical galaxies. Schade et al.\cite{s3}$^,$\cite{s4}
analysed ground-based and $HST$ imaging
and find an increase with redshift in surface 
brightness or luminosity (at a given size)
of $\Delta M_B \sim -z$, consistent with both the spectroscopic
studies cited above and with other imaging work\cite{pa}$^,$\cite{ba}.  
 Thus, the luminosity evolution expected from passive evolution models
of elliptical galaxies has been detected by several groups.

\newpage

%\vspace{-1.5cm}\vbox{
\centerline{ }
\vspace{1.3cm}\vbox{
\hbox{
\hbox{
\includegraphics{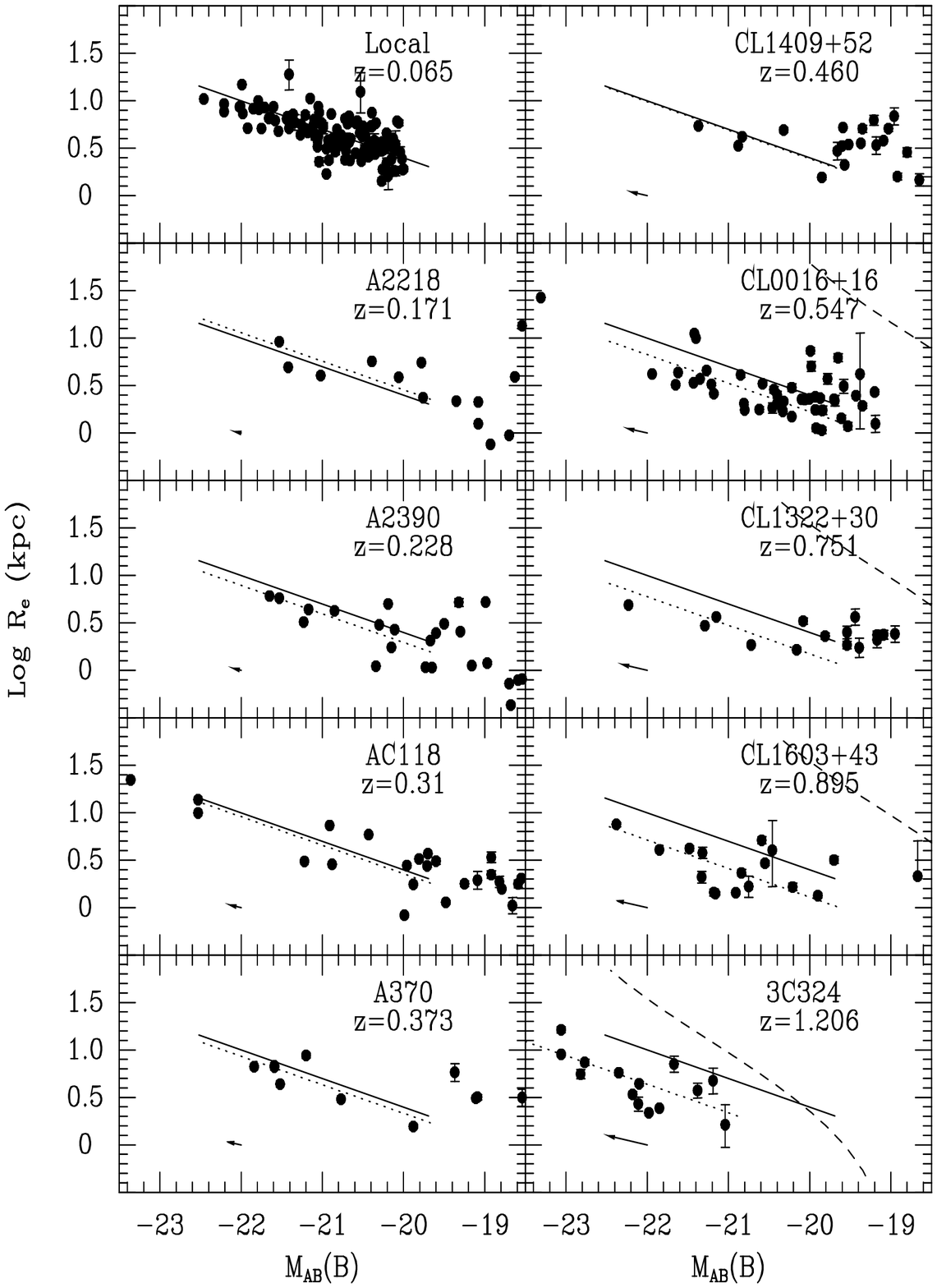} }
\qquad
\qquad
\hbox{ \hfill
\hspace{-2.0cm}\vspace{3.0cm}
\includegraphics{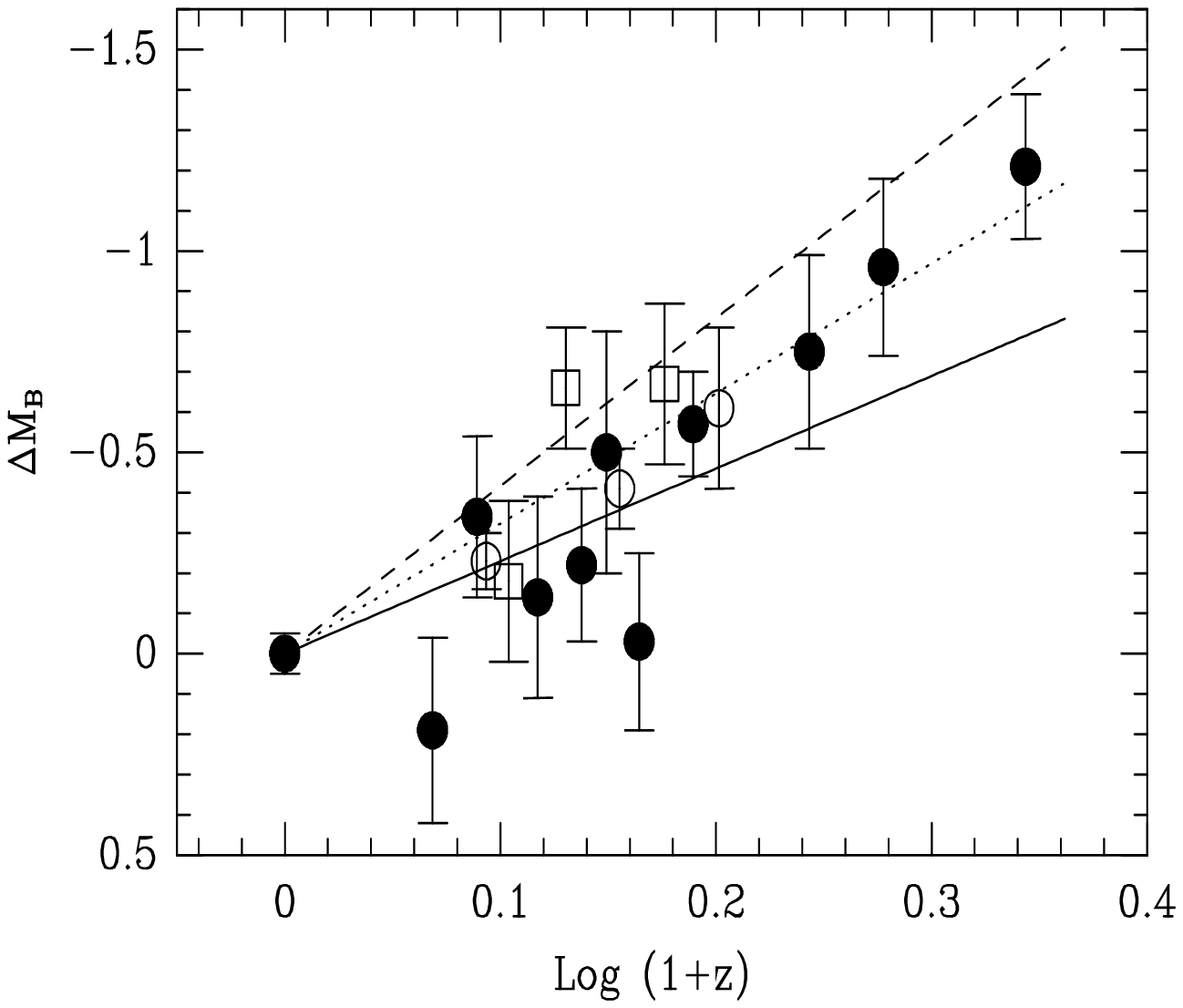} }
%\special{psfile=SB4pg.ps voffset=-430 hoffset=170 vscale=50 hscale=60 angle=0}
}
 }

\vspace{6.5cm}
\hbox{
\hspace{-0.5cm}\hbox{\vspace{-1.0cm}\parbox{6.0truecm}{\scriptsize\baselineskip
8pt
{\bf Figure 1:} The evolving size-luminosity relation for elliptical galaxies
derived from $HST$ archival imaging of 9 clusters with  $0 < z < 1.2$\cite{s4}.
Galaxies
of a given size are more luminous by $\sim 1$ mag at $z=1$. Solid lines show the
local relation  and dotted lines indicate the evolved relation measured in each
cluster. (Long-dashed lines indicate the surface-brightness selection.)
}}

\qquad
\qquad
\qquad
\vspace{-1.0cm}\hspace{-1.5cm}\hbox{\parbox{5.5truecm}{\scriptsize\baselineskip
8pt
{\bf Figure 2:} $\Delta M_B$ is the shift in luminosity at a given size as
measured, e.g., from figure 1. Solid symbols are for cluster elliptical galaxies
using $HST$ imaging (but no membership information) and open symbols are from
ground-based imaging of CNOC fields\cite{s3} 
where all of the galaxies have
redshift information (open circles=cluster E's,open squares=field E's.)

%here

 } %end of caption hbox
 } %added
 } %added

% } % end of vbox

\vspace{0.75cm}

\subsection{Blue cluster galaxies}

 The blue fraction of the cluster galaxy population increases
from a few percent locally to $25\%$ at $z\sim 0.5$\cite{bu} 
and, by $z\sim 0.9$, perhaps
80\% of cluster galaxies were blue.\cite{ra} 
 This blue population shows spectroscopic signs of enhanced
levels of star-formation\cite{co} and
is made up largely of disk-like galaxies with a high frequency
of peculiar/irregular 
structure.\cite{th}$^,$\cite{la}$^,$\cite{dr}$^,$\cite{co2}
 Surface photometry of galaxies (with redshifts) from the Canadian Network
for Observational Cosmology (CNOC) cluster survey\cite{ca} 
shows that
galactic disks in 3 clusters have surface brightness higher
than the Freeman\cite{fr} value by $\sim 1$ mag at $z=0.55$\cite{s2}.
Furthermore, this disk brightening is consistent with 
observations of field galaxies.\cite{s1}$^,$\cite{s95}$^,$\cite{s2}

\newpage
\vspace{0.0cm}\vbox{
\hbox{
\hbox{
\includegraphics{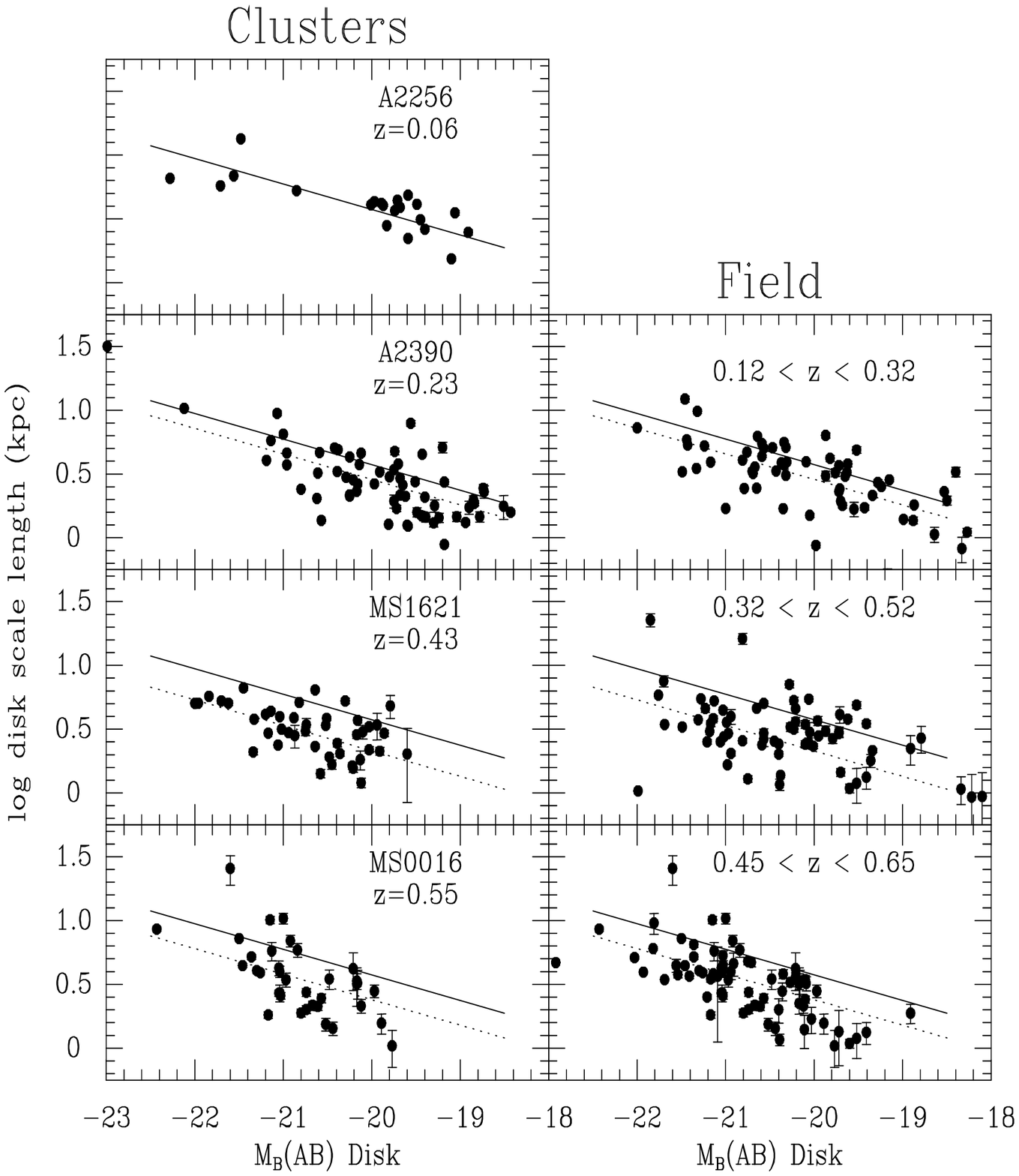} }
\qquad
\qquad
\hbox{ \hfill
\hspace{-2.0cm}\vspace{3.0cm}
\includegraphics{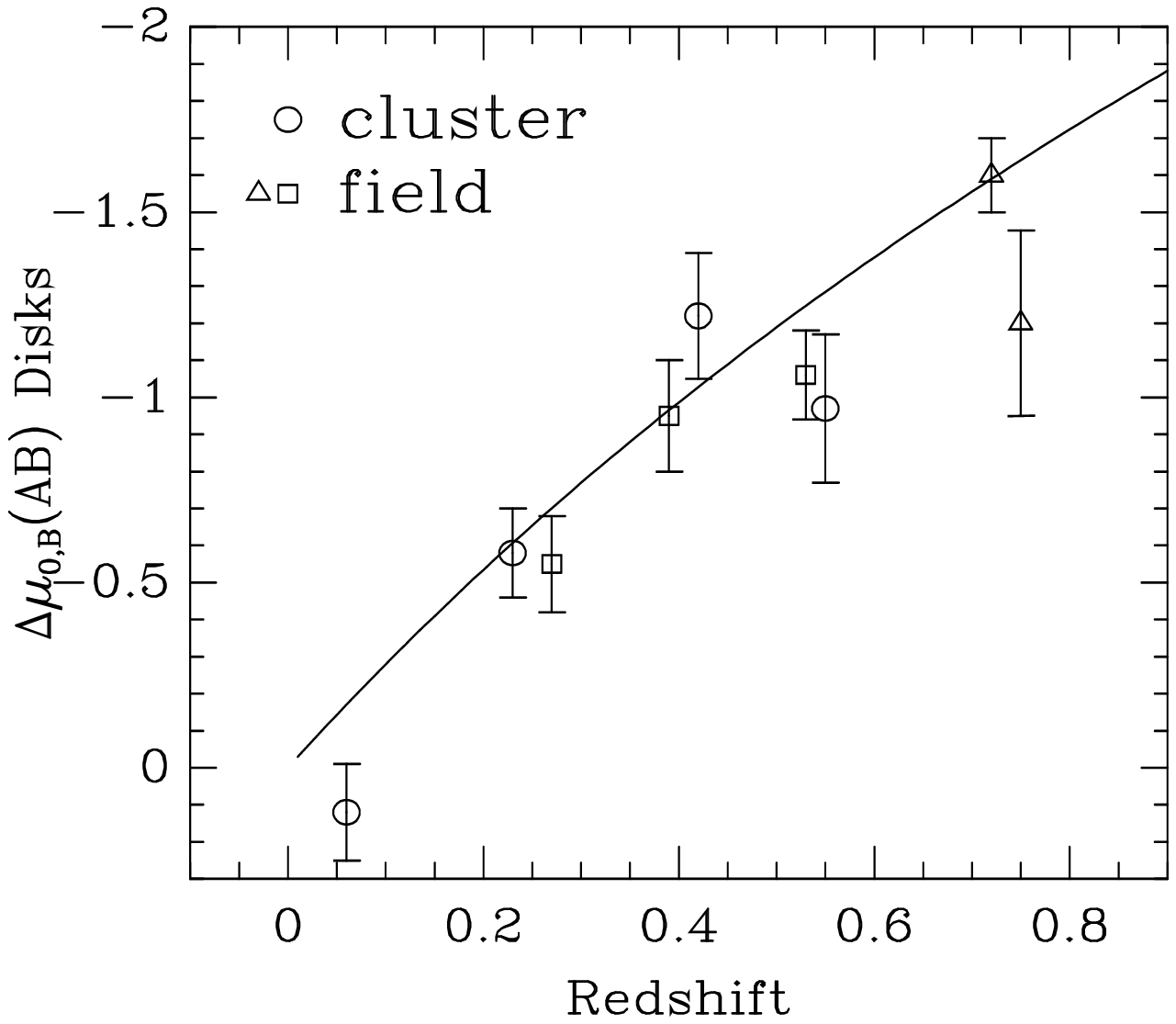} }

}
 }

\vspace{6.5cm}
\hbox{
\hspace{-0.5cm}\hbox{\vspace{-1.0cm}\parbox{6.0truecm}{\scriptsize\baselineskip 8pt
{\bf Figure 3:} The evolving size-luminosity relation for galactic disks 
in clusters and the field\cite{s2}.
Solid lines show the Freeman\cite{fr} law
and dotted lines indicate the mean evolved relation.

}}

\qquad
\qquad
\qquad
\vspace{-0.0cm}\hspace{-1.5cm}\hbox{\hfill\parbox{5.5truecm}{\scriptsize\baselineskip
8pt
{\bf Figure 4:} $\Delta M_B$ is the shift in surface brightness
relative to the Freeman law measured, e.g. from plots like Figure 3.
Open circles are CNOC cluster galaxies and squares are CNOC field
galaxies. Triangles are field galaxies from 
Schade et al.\cite{s95}$^,$\cite{s2}.
The line is the evolution of the luminosity density ($(1+z)^{2.7}$)
in the $B$-band from Lilly et al.\cite{li}.

%here

 } %end of caption hbox
 } %added
 } %added

% } % end of vbox

\vspace{-0.4cm}

\section{Parallels of cluster and field population at high-z}

\vspace{-0.3cm}
Cluster and field populations are dominated at high redshift
by a ``blue-excess'' population that is mysteriously ``absent''
from the local population\cite{li2}$^,$\cite{dr}.  
Both populations show a high (and similar)
frequency of peculiar/irregular structure,\cite{gla}$^,$\cite{co2} 
and both populations
show elevated rates of star formation relative to local
populations\cite{ha}$^,$\cite{co}. It has been argued 
here that field and cluster
populations also show evolution in surface brightness 
(both disks and ellipticals) and that the rates of
evolution are not measurably different (although the
cluster population we have studied consists largely of
galaxies far from the dense cluster core). 

 These similarities suggest that much of the physics of
galaxy evolution is common to cluster and field populations
and that the development of the high-z cluster population, largely
through infall of field galaxies, needs to be understood
in the context of the field population {\em at that redshift}, 
whose state differs markedly from the population we
see in the present-day universe.

\nopagebreak
\vspace{-0.7cm}
\section*{Acknowledgments}
\vspace{-0.3cm}
 It is a pleasure to thank the many collaborators 
in the Canada-France Redshift Survey and the Canadian
Network for Observational Cosmology groups, in particular
Simon Lilly, Ray Carlberg, and Felipe Barrientos.

\section*{References}

%
% The following journals are predefined in the .sty file:
%
% \def\aj{{AJ}}			
% \def\araa{{ARA\&A}}		
% \def\apj{{ApJ}}			
% \def\apjl{{ApJ}}		
% \def\apjs{{ApJS}}		
% \def\ao{{Appl.~Opt.}}		
% \def\apss{{Ap\&SS}}		
% \def\aap{{A\&A}}		
% \def\aapr{{A\&A~Rev.}}		
% \def\aaps{{A\&AS}}		
% \def\azh{{AZh}}			
% \def\baas{{BAAS}}		
% \def\jrasc{{JRASC}}		
% \def\memras{{MmRAS}}		
% \def\mnras{{MNRAS}}		
% \def\pra{{Phys.~Rev.~A}}	
% \def\prb{{Phys.~Rev.~B}}	
% \def\prc{{Phys.~Rev.~C}}	
% \def\prd{{Phys.~Rev.~D}}	
% \def\pre{{Phys.~Rev.~E}}	
% \def\prl{{Phys.~Rev.~Lett.}}	
% \def\pasp{{PASP}}		
% \def\pasj{{PASJ}}		
% \def\qjras{{QJRAS}}		
% \def\skytel{{S\&T}}		
% \def\solphys{{Sol.~Phys.}}	
% \def\sovast{{Soviet~Ast.}}	
% \def\ssr{{Space~Sci.~Rev.}}	
% \def\zap{{ZAp}}			
% \def\nat{{Nature}}		
% \def\iaucirc{{IAU~Circ.}}
% \def\aplett{{Astrophys.~Lett.}}
% \def\apspr{{Astrophys.~Space~Phys.~Res.}}
% \def\bain{{Bull.~Astron.~Inst.~Netherlands}}
% \def\fcp{{Fund.~Cosmic~Phys.}}
% \def\gca{{Geochim.~Cosmochim.~Acta}}
% \def\grl{{Geophys.~Res.~Lett.}}
% \def\jcp{{J.~Chem.~Phys.}}	
% \def\jgr{{J.~Geophys.~Res.}}	
% \def\jqsrt{{J.~Quant.~Spec.~Radiat.~Transf.}}
% \def\memsai{{Mem.~Soc.~Astron.~Italiana}}
% \def\nphysa{{Nucl.~Phys.~A}}
% \def\physrep{{Phys.~Rep.}}
% \def\physscr{{Phys.~Scr}}
% \def\planss{{Planet.~Space~Sci.}}
% \def\procspie{{Proc.~SPIE}}
%

\end{document}